\documentclass[10pt,journal,letterpaper,compsoc]{IEEEtran}
\usepackage{graphics}
\usepackage{epsfig}
\usepackage{graphicx}
\graphicspath{{./figures/}}
\usepackage{url}

\begin{document}
%
\title{Precise, Scalable and Online Request Tracing for Multi-tier Services of Black Boxes}\label{introduction}
%
%
%
%

\author{Bo~Sang,
        Jianfeng~Zhan,
        Zhihong~Zhang,
        Lei~Wang,
        Dongyan ~Xu,
        Yabing~Huang and
        Dan~Meng
\IEEEcompsocitemizethanks{\IEEEcompsocthanksitem Bo Sang, Jianfeng Zhan, Zhihong Zhang, Lei Wang, Yabing Huang and Dan Meng are with the Institute of Computing Technology, Chinese Academy of Sciences. Dongyan Xu is with the Department of Computer Science, Purdure University. Jianfeng Zhan is the corresponding author. E-mail: jfzhan@ncic.ac.cn. \protect\\
}
}

\IEEEcompsoctitleabstractindextext{%
\begin{abstract}
As more and more multi-tier services are developed from commercial off-the-shelf components or heterogeneous middleware \emph{without source code available}, both developers and administrators need a request tracing tool to (1) exactly know how a user request \emph{of interest} travels through services of \emph{black boxes}; (2) obtain macro-level user request behavior information of services without the necessity of inundating within massive logs. This need is further exacerbated by the IT system "agility"\cite{vPath}, which mandates the tracing tool to \emph{on-line} offer performance data since off-line approaches can not reflect system changes in real time. Moreover, taking it into account the large scale of deployed services, a pragmatic tracing approach should be scalable in terms of the cost in collecting and analyzing logs.

Previous research efforts either accept imprecision of probabilistic correlation methods or present precise but \emph{unscalable} tracing approaches that have to collect and analyze large amount of logs; Besides, previous precise request tracing approaches of black boxes fail to propose macro-level abstractions that enables debugging performance-in-the-large, and hence users have to manually interpret massive logs. This paper introduces a \emph{precise}, \emph{scalable} and \emph{online} request tracing tool, named \emph{PreciseTracer}, for multi-tier services of black boxes. Our contributions are four-fold: first, we propose a precise request tracing algorithm for multi-tier services of black boxes, which only uses application-independent knowledge; second, we respectively present micro-level and macro-level abstractions: \emph{component activity graphs} and \emph{dominated causal path patterns} to represent \emph{causal paths of each individual request} and \emph{repeatedly executed causal paths that account for significant fractions}; third, we present two mechanisms: \emph{tracing on demand} and \emph{sampling} to significantly increase system scalability; fourth, we design and implement an online request tracing tool. \emph{PreciseTracer}'s fast response, low overhead and scalability make it a promising tracing tool for large-scale production systems.
\end{abstract}

}

\maketitle

\IEEEdisplaynotcompsoctitleabstractindextext

%
\IEEEpeerreviewmaketitle

\section{Introduction} \footnote{This is an extended work of our paper: Z. Zhang, J. Zhan, Y. Li, L. Wang and D. Meng, Precise request tracing and performance debugging for multi-tier services of black boxes, The 39th IEEE/IFIP International Conference on Dependable Systems \& Networks (DSN '09), 2009.}
%
%

%
%
%
%
\IEEEPARstart{A}{s} more and more multi-tier services are developed from commercial off-the-shelf components or heterogeneous middleware \emph{without source code available}, both developers and administrators need a request tracing tool to (1) exactly know how a user request \emph{of interest} travels through services of \emph{black boxes} if necessary; (2) obtain macro-level performance information of services without the necessity of inundating within massive logs. This need is further exacerbated by IT system ¡°agility¡±\cite{vPath}. For example, in data centers, service customers increasingly require support for peak loads that are an order of magnitude greater than those experienced in normal steady state \cite{Oceano}. To meet with fluctuated workloads, resources are often dynamically provisioned, and hence service instances are adjusted accordingly. In this context, system agility mandates tracing tools to \emph{on-line} offer performance information, since off-line tracing approaches can not reflect system changes in real time.

There are several challenges in proposing a request tracing tool for on-line analysis of multi-tier services of black boxes. First of all, those tools should not produce disturbed impact on performance of multi-tier services. Second, services are often deployed within data centers with a large scale, and hence a pragmatic tracing tool should be scalable in terms of the cost in collecting and analyzing logs\cite{Dapper}. Finally, administrators not only needs to exactly track a user request \emph{of interest} if necessary, but also needs to focus on \emph{performance-in-the-large} \cite{PerformanceDebug}. A precise request tracing system will inevitably produce massive logs, however debugging \emph{inundated with massive logs} is not a trivial issue, and hence we need to propose macro-level abstractions to facilitate debugging performance-in-the-large of multi-tier services.

The most straightforward and accurate way \cite{Dapper}\cite{ETE} \cite{Magpie} \cite{NetLogger} \cite{Pinpoint}\cite{Stardust} to correlate message streams is to leverage application-specific knowledge and explicitly declare causal relationships among events of different components. The disadvantage is that user must obtain and modify source code of target applications or middleware, or even require that users have in-depth knowledge of target applications or instrumented middleware, and hence they cannot be used for services of black boxes. Without the knowledge of source code, several previous approaches \cite{PerformanceDebug}\cite{Wap5}\cite{E2EProf} either accept \emph{imprecision of probabilistic correlation methods} to infer average response time of components\cite{vPath}, or \cite{BorderPatrol} rely upon the knowledge of protocols to isolate events or requests for precise request tracing. Probably closest to our work is \emph{vPath} \cite{vPath}, which presents a precise but \emph{unscalable} request tracing tool for services of black boxes. Because of its implementation limitation, the tracing mechanism in \emph{vPath} can not be enabled or disabled on demand \emph{without interrupting services}, and hence it has to continuously collect and analyze logs, which results in unacceptable cost. For example, as stated in \cite{Magpie2},  a simple e-commercial system would generates $10 M$ log data per minute. Usually, a data center has $10$ thousands or even more nodes being deployed with multi-tier services. If we debug performance problems of multi-tier services on this scale for one minute, a tracing system with \emph{continuous logging} needs to analyze at least $0.1$ $TB$ logs, which in nature is unscalable. Moreover, state-of-the-art precise requesting tracing approaches of black boxes \cite{BorderPatrol} \cite{vPath} fail to propose abstractions for representing macro-level user request behaviors, and instead depend on users' manual interpretation of logs in debugging performance-in-the-large. Besides, they \cite{BorderPatrol} \cite{vPath} are offline.

In this paper, we present a \emph{precise} and \emph{scalable} request tracing tool for online analysis of multi-tier services of black-boxes. Our tool collects activity logs of multi-tier services through kernel instrumentation, which can be enabled or disabled on demand. Through tolerating loss of logs, our system supports \emph{sampling} or \emph{tracing on demand}, which significantly decreases the collected and analyzed logs and hence improves the system scalability. After reconstructing those activity logs into \textit{causal paths}, each of which is  \emph{a sequence of activities with causal relations caused by an individual request}, we classify those \textit{causal paths} into different \emph{patterns} according to their shapes, and present an macro-level abstraction, \emph{dominated causal path patterns}, to represent \emph{repeatedly executed causal paths that account for significant fractions}. Finally, we detect performance problems through observing changes of performance data of \emph{dominated causal path patterns}. In this paper, we make the following contributions:
 \begin{enumerate}
   \item We design a precise tracing algorithm to deduce causal paths of requests from interaction activities of components of black boxes. Our algorithm only uses application-independent knowledge.
   \item We present two abstractions: \emph{component activity graphs} and \emph{dominated causal path patterns} to respectively represent micro-level and macro-level user request behaviors of services.
   \item We present two mechanisms: \emph{tracing on demand} and \emph{sampling}, to improve system scalability. Our experiments show \emph{sampling} decreases the cost of collecting and analyzing logs, and at the same time still preserves performance data of services in the way that it captures most of \emph{dominated causal path patterns}.
   \item We design and implement an online request tracing system. Our experiments demonstrate that the fast response, low overhead and robustness of \emph{PreciseTracer} make it a promising tracing tool for large-scale production systems.
       \end{enumerate}

The paper is organized as follows: Section \ref{problem_statement} formulates the problem; Section \ref{related_work} summarizes related work; Section \ref{design_implementation} describes \emph{PreciseTracer} design and implementation; Section \ref{evaluation} evaluates our tool.  Section \ref{conclusion} draws a conclusion.

\section{Problem statement} \label{problem_statement}

Our target environments are data centers deployed with multi-tier services. There are two types of nodes in this environment: \emph{service nodes} and \emph{analysis nodes}. \emph{Service nodes} are the ones on which multi-tier services are deployed, while most component of the tracing tool are deployed on \emph{analysis nodes}.

\begin{figure}[hbtp]
  \centering
  \includegraphics[scale=0.45]{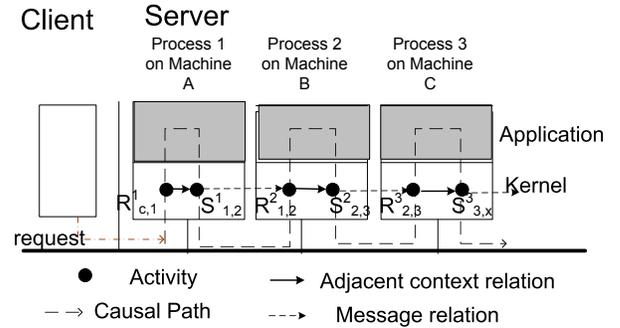}\\
  \caption{Activities with causal relations in the kernel}
  \label{request_observation}
\end{figure}

We treat each component in a multi-tiers service as a black box. That is to say we can not obtain the application or middleware source code. In Fig. \ref{request_observation}, we observe that a request will cause a series of \emph{interaction activities in the operating system kernel} , e.g. sending or receiving messages. Those activities happen under specific contexts (\emph{processes} or\emph{ kernel threads}) of different components. We record an activity of sending a message as $S^{i}_{i,j}$, which indicates a process $i$ sends a message to a process $j$. We record an activity of receiving a message as $R^{j}_{i,j}$, which indicates a process $j$ receives a message from a process $i$.

When an individual request is serviced, a series of activities \emph{having causal relations or happened-before relationships as defined by Lamport \cite{Lamport}} constitute \emph{a causal path}. For example in Fig. \ref{request_observation}, the activities sequence \{$R^{1}_{c,1}, S^{1}_{1,2}, R^{2}_{1,2}, S^{2}_{2,3}, R^{3}_{2,3}, S^{3}_{3,x}$\} constitutes \emph{a causal path}. \emph{For each individual request, there is a causal path}.

Our project develops a tracing tool to help developers and administrators in the following ways:
\begin{enumerate}
  \item Precisely correlate activities of components \emph{caused by each request} into individual causal paths, through which user can exactly track how a user request \emph{of interest} travels through services. \label{item_1}
  \item Identify \emph{dominated causal path patterns}, and obtain their statistic information, e.g. the average service time of each component in different patterns. \label{item_2}
  \item Debug performance problems of a multi-tier service with the help of \ref{item_1}) and \ref{item_2}).
  \item Provide online performance data of services for the feedback controller in the runtime power management system to save cluster power consumption, which is resolved in our another research project\cite{PowerTracer}.
\end{enumerate}

The application limits of our system are as follows:
\begin{enumerate}
  \item We do not require source code of applications, neither deploy the instrumented middleware, and neither have the knowledge of high-level protocols used by services, like http etc.
  \item A single execution entity (\emph{a process} or \emph{a kernel thread}) of each component can only service one request in a certain period. For serving each individual request, execution entities of components cooperate through sending or receiving messages using a reliable communication protocol, like TCP.
\end{enumerate}

Though not all multi-tier services fall within the scope of our target applications, fortunately many popular services satisfy these assumptions. For example, our method can be used to analyze concurrent servers following nine design patterns introduced in \cite{UnixNetwork}, including iteration model, concurrent process model and concurrent thread model.

\section{Related work} \label{related_work}

\begin{table}[hbtp]
\renewcommand{\arraystretch}{1.3}
\caption{Comparisons of black-box tracing approaches}
\label{relevant_work_comparison}
\centering
\begin{tabular}{|l|p{1.0cm}|p{1.6cm}|p{1.5cm}|p{0.7cm}|}
  \hline
  & \itshape accuracy & \itshape scalability & \itshape request abstraction & \itshape mode \\ \hline

  \itshape Project5 &imprecise &/ &macro level &offline \\ \hline
  \itshape E2EProf  & imprecise &/ &macro level &online \\ \hline
  \itshape BorderPatrol  &precise &continuous logging &micro level &offline \\ \hline
  \itshape vPath  &precise &continuous logging &micro level &offline \\ \hline
  \itshape Our tool  &precise &sampling \& tracing on demand &micro/macro levels &online \\ \hline
  \hline
\end{tabular}
\end{table}

\textbf{Imprecise black-box approaches}. A much earlier project, \emph{DPM} \cite{DPM} instruments the operating system kernel and tracks the causality between pairs of message to trace \emph{unmodified applications}. \emph{DPM} is not precise, since the existence of a path in resulting graphs does not necessarily mean that any real causal path followed all of those edges in that sequence \cite{BorderPatrol}. \emph{Project5} \cite{PerformanceDebug} and \emph{WAP5} \cite{Wap5} accept  imprecision of probabilistic correlations. \emph{Project5} proposes two algorithms intended for \emph{offline analysis}: a nesting algorithm assumes 'RPC-style' (call-returns) communication, while a convolution algorithm does not assure a particular messaging protocol. The nesting algorithm only uses one timestamp per message\cite{Wap5} without distinguishing SEND or RECEIVE timestamp, and hence it only provides aggregate information per component. The convolution algorithm only infers average response time of components, and can not build individual causal paths (micro-level request abstraction) for each request. More recently \emph{WAP5} \cite{Wap5} infers causal paths for wide-area systems from tracing stream on a per-process granularity using library interposition. Similar to the convolution algorithm of \emph{Project5}, \emph{E2Eprof} \cite{E2EProf} proposes a pathmap algorithm, and uses compact trace representations and a series of optimizations make it suitable for online performance diagnosis. As claimed by themselves \cite{E2EProf}, \emph{E2EProf} only tolerate small clock skews (i.e., equal to few times of the time quanta in term of milliseconds), when determining service paths. If the skew is large, analysis results will not be accurate. Adopting probabilistic correlation in those approaches \cite{PerformanceDebug}\cite{Wap5}\cite{E2EProf} should facilitate reducing the cost in collecting and analyzing logs, though the system scalability is not demonstrated in \cite{PerformanceDebug} \cite{E2EProf} \cite{Wap5}.

\textbf{Precise black-box approaches}. There are only two precise black-box approaches: \emph{BorderPatrol} and \emph{vPath}, which are \emph{offline} and can not offer performance information in real time. With the knowledge of diverse protocols used by multi-tier service, \emph{BorderPatrol} isolates and schedules events or requests at the protocol level to precisely trace requests. When multi-tier services are developed from commercial components or heterogeneous middleware, \emph{BorderPatrol} needs to write many protocol processors and requires more specialized knowledge than pure black-box approach \cite{BorderPatrol}. \emph{VPath} consists of an monitor and an offline log analyzer. The  monitor continuously logs which thread performs a send or recv system call over which TCP connection. The offline log analyzer parses logs generated by the monitor to discover request-processing paths. The monitor is implemented in virtual machine monitor (VMM) through system call interceptions. Using library interposition (BorderPatrol) or system call interception(vPath) , the logging mechanism of \emph{BorderPatrol} or \emph{vPath} can not be enabled or disabled on demand without interrupting services, and hence it is unscalable in terms of the cost of collecting and analyzing logs. Beside, \emph{BorderPatrol} and \emph{vPath} fails to present macro-level abstractions to facilitate debugging performance-in-the-large, and users have to inundate within massive logs with great efforts.

The most invasive systems, such as \emph{Netlogger} \cite{NetLogger} and \emph{ETE} \cite{ETE} require programmers to add event logging to carefully-chosen points to find causal paths rather than infer them from passive traces. \emph{Pip} \cite{Pip} inserts annotations into source code to log actual system behaviors, but can extract causal path information with no false positives or false negatives. \emph{Magpie} \cite{Magpie} collects events at different points in a system and uses an event schema to correlate these events into causal paths. In order to track a request from end to end, magpie must obtain the source code of application, at least require "wrapper" around some part of the application \cite{PerformanceDebug}. \emph{Stardust} \cite{Stardust}, a system used as an on-line monitoring tool in a distributed storage system, is implemented in a similar way. \emph{Whodunit} \cite{Whodunit} annotates profile data with transaction context synopsis, and tracks and profiles transactions that flow through shared memory, events, or via inter- process communication using messages, which provides finer grained knowledge of transactions and their profile data within each box.

To avoid modifying applications' source code, several previous efforts enforce middleware or infrastructure change, bound to specific middleware or deployed instrumented infrastructure. \emph{Pinpoint} \cite{Pinpoint} locates component faults in J2EE platforms by tagging and propagating a globally unique request ID with each request. \emph{Causeway} \cite{Causeway} enforces change to network protocol so as to tag meta-data with existing module communication. \emph{X-Trace} \cite{X-Trace} modifies each network layer to carry X-Trace meta-data that enables path casual path reconstruction and focuses on debugging paths through many network layers. The recent Google technical report shows that its production tracing infrastructure \emph{Dapper} uses a global identifier to tie together related events from various parts of a distributed system. \emph{Dapper} uses sampling to improve system scalability and reduce performance overhead. With the support of logging mechanism of Hadoop, J. Tan et al. \cite{Causal_Tracing_MapReduce} presented a non-intrusive approach to tracing the causality of execution in MapReduce systems, which is significantly different from multi-tier services.

\section{PreciseTracer Design and Implementation} \label{design_implementation}
Section \ref{architecture}, \ref{section_tracing_algorithm}, \ref{section_scalability} and Section \ref{implementation} respectively presents \emph{PreciseTracer} architecture, tracing algorithm, system scalability issue and system implementation.
\subsection{PreciseTracer Architecture} \label{architecture}
\begin{figure}[hbtp]
  \centering
  \includegraphics[scale=0.45]{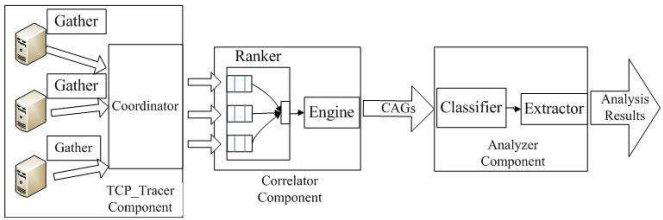}\\
  \caption{PreciseTracer Architecture}
  \label{precisetracer_architecture}
\end{figure}

\emph{PreciseTracer} is flexible, and administrators could configure the system according to their requirements. \emph{PreciseTracer} could work under \emph{a offline model} or  \emph{an online model}. Under an offline model, \emph{PreciseTracer} only analyze logs after collecting logs for the specified period, which is often quite a long time. Under an online model, \emph{PreciseTracer} will collect and analyze logs simultaneously, providing administrators with real-time performance information.

Fig \ref{precisetracer_architecture} shows \emph{PreciseTracer} architecture, including three major components: \emph{TCP\_Tracer}, \emph{Correlator} and \emph{Analyzer}. \emph{TCP\_Tracer} logs all interaction activities of components, and provides \emph{Correlator} with logs of multi-tier service. \emph{Correlator} is responsible for correlating those activity logs of different components into \emph{causal paths}. Finally based on causal paths produced by \emph{Correlator}, \emph{Analyzer} extracts useful information and outputs analysis results.

\subsubsection{TCP\_Tracer} \label{TCP_Tracer}
\emph{TCP\_Tracer} includes two modules: \emph{Gather} and \emph{Coordinator}. Deployed on each \emph{service node}, \emph{Gather} is responsible for collecting logs of services deployed on the same node. \emph{Coordinator}, which is deployed on \emph{a analysis node}, is in charge of controlling \emph{Gather modules} on each \emph{service node}. \emph{Coordinator} configures and synchronies \emph{Gather modules} to send logs to \emph{Correlator} deployed on \emph{a analysis node}.

\emph{TCP\_Tracer} independently observe interaction activities of components of black boxes on different nodes. Concentrating on the main focus, \emph{TCP\_Tracer} only concern about when serving a request starts, finishes, and when components receive or send messages \emph{within the confine of a data center}. Of course, we can extend our concerns to observe more activities if the overhead is acceptable. In this paper, our concerned activities types include: \emph{BEGIN}, \emph{END}, \emph{SEND}, and \emph{RECEIVE}. \emph{SEND} and \emph{RECEIVE} activities are the ones of sending and receiving messages. A \emph{BEGIN} activity marks the start of serving a new request, while an \emph{END} activity marks the end of servicing a request.

For each activity, \emph{TCP\_Tracer} log five attributes: \emph{activity type, timestamp, context identifier, message identifier} and \emph{message size}. For each activity, we use (\emph{hostname, program name, process ID, thread ID}) tuple to describe its context identifier, and use (\emph{IP of sender, port of sender, IP of receiver, port of receiver, message size}) tuple to describe its message identifier.

\subsubsection{Correlator}
\emph{Correlator} includes two major modules: \emph{Ranker} and \emph{Engine}. \emph{Ranker} is responsible for choosing candidate activities for composing causal paths; \emph{Engine} constructs causal paths from the outputs of \emph{Ranker}, and then outputs them.

Formally, a \emph{causal path} can be described as \emph{a directed acyclic graph} ${G(V, E)}$, where vertices $V$ are activities set of components, and edges $E$ represent causal relations between activities. We define this abstraction as \emph{component activity graph} (\emph{CAG}). For an individual request, a corresponding \emph{CAG} represents all activities with causal relations in the life cycle of serving an individual request.

CAGs include two types of relations: \emph{adjacent context relation} and \emph{message relation}. We formally define two relations based on the happened-before relation \cite{Lamport}, which is denoted as $\rightarrow$, as follows:

\emph{Adjacent Context Relation}: caused by the same request $r$, $x$ and $y$ are activities observed in the same context $c$ (\emph{a process} or \emph{a kernel thread}), and $x \rightarrow y$ holds true. If no activity $z$ \emph{that satisfies the relations $x \rightarrow z$ and $z \rightarrow y$} is observed in the same context, we can say \emph{an adjacent context relation} exits between $x$ and $y$, denoted as $x \rightarrow_{c} y$. So \emph{the adjacent context relation} $x \rightarrow_{c} y$ means that $x$ has happened right before $y$ in the same execution entity.

\emph{Message Relation}: for serving a request $r$, if $x$ is a \emph{SEND} activity, which sends a message $m$, and $y$ is a \emph{RECEIVE} activity, which receives the same message $m$, then we can say \emph{a message relation} exists between $x$ and $y$, denoted as $x \rightarrow_{m} y$. So \emph{the message relation} $x \rightarrow_{m} y$ means that $x$, which sends a message, has happened right before $y$, which receives a message, in two different execution entities.

If there is an edge from activity $x$ to activity $y$ in a CAG, for which $x \rightarrow_{c} y$ or $x \rightarrow_{m} y$ holds true, then $x$ is the parent of $y$.

In a CAG, every activity vertex must satisfy the property: \emph{each activity vertex has no more than two parents, and only a RECEIVE activity vertex could have two parents, with which one parent has an adjacent context relation and another one has a message relation}.

\begin{figure}[hbtp]
  \centering
  \includegraphics[scale=0.44]{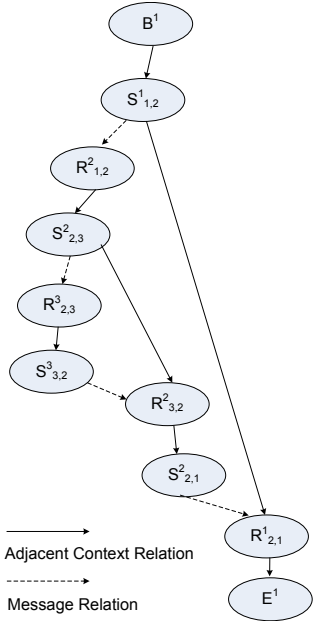}\\
  \caption{An example of an individual CAG}
  \label{CAG_example}
\end{figure}

For an individual request, it is clear that correlating a causal path is the course of building a CAG with interaction activities as the input. Fig.\ref{CAG_example} shows an example of an individual CAG.

\subsubsection{Analyzer} \label{section_analyzer}

\emph{Analyzer} includes two major modules: \emph{classifier} and \emph{extractor}. \emph{Classifier} is responsible for classifying \emph{causal paths} into different \emph{causal path patterns}, while \emph{Extractor} provides analysis results based on \emph{causal path patterns}.

CAGs include rich performance data of services, because a CAG indicates how a request from clients is served by each component. For example, if administrators want to pinpoint the performance bottleneck of a service, they need to obtain the service time consumed on each component in serving requests. According to a CAG, we can calculate latencies of components in serving an individual request. For example, for the request in Fig. \ref{request_observation}, the latency of \emph{process 2} is $(t(S_{2, 3}) - t(R_{1, 2}))$, and the interaction latency from \emph{process 1} to \emph{process 2} is $(t(R_{1, 2}) - t(S_{1, 2}))$, where $t$ is the local timestamp of each activity. The latency of \emph{process 2} is accurate, since all timestamps are from the same node. The interaction latency from \emph{process 1} to \emph{process 2} is inaccurate, since we do not remedy the clock skew between two nodes.

A single causal path could help administrators get \emph{micro-level user request information} of services. For example, administrators could detect failed nodes if some causal paths show abnormal information. However, causal paths could not be directly utilized to represent macro-level performance data of services directly for two reasons: (1) There are massively causal paths. An individual causal path could only reflect how a request was served by services. To consider the disturbances in environment, it's not so credible to take an individual causal path as service's performance data; (2) different types of requests would produce causal paths with different features. Thus, we need a \emph{macro-level abstraction} to represent performance data of multi-tier services.

We can classify \emph{causal paths} into different \emph{causal path patterns} according to shapes of CAGs, and further figure out which ones are dominated causal path patterns according to their fractions in terms of path number. CAGs could be classified into the same causal path pattern when they meets with the following criteria:
\begin{enumerate}
  \item There are the same number of activities in two CAGs;
  \item Two matching activities with the same order in two respective CAGs have the same attribution in terms of \emph{(activity type, program name)}.
\end{enumerate}

In a CAG, for each activity, we define its order according to the following rules:

\textbf{Rule 1}: If $x \rightarrow_{c} y$ or $x \rightarrow_{m} y$, then $x \prec y$;

\textbf{Rule 2}: If $x \rightarrow_{c} y$ and $x \rightarrow_{m} z$ and there is no relation between $y$ and $z$, then $x \prec z \prec y$;

\textbf{Rule 3}: If $y \rightarrow_{c} x$ and $z \rightarrow_{m} x$ and there is no relation between $y$ and $z$, then $y \prec z \prec x$.


After the classification, we could compute the average performance data about \emph{dominated causal path patterns}, on a basis of which we could further detect performance problems of multi-tier service. Our experiments in Section \ref{injected_problems} demonstrate the effectiveness of this approach.

\subsection{The tracing algorithm} \label{section_tracing_algorithm}
Before we proceed to introduce the algorithm of \emph{ranker}, we explain how \emph{engine} stores \emph{incomplete CAGs}. In the course of building CAGS, all incomplete CAGs are indexed with two \emph{index map} data structures. \emph{An index map} maps a key to a value, and supports basic operations, like search, insertion and deletion. One index map, named \emph{mmap}, is used to match \emph{message relations}, and another one, named \emph{cmap}, is used to match \emph{adjacent context relations}. For \emph{mmap}, the key is \emph{the message identifier of an activity}, and the value of \emph{mmap} is \emph{an unmatched SEND activity with the same message identifier}. The key in \emph{cmap} is  \emph{the context identifier of an activity}, and the value of \emph{cmap} is the \emph{latest activity with the same context identifier }.

Section \ref{section_choose_activity} explains how to choose candidate activities in constructing CAGs in our algorithm; Section \ref{section_construct_cags} introduces how to construct CAGs; and Section \ref{section_disturbance_tolerance} describes how to handle disturbances.

\subsubsection{Choosing candidate activities for composting CAGs} \label{section_choose_activity}
For each \emph{service node}, we choose the minimal local timestamp of activities as the initial time. We set a \emph{sliding time window} for processing activity stream, and activities, logged on different nodes, will be fetched into the buffer of \emph{Ranker} if their timestamps are within the sliding time window. Section \ref{section_disturbance_tolerance} will present to how to deal with clock skews in distributed systems.

\emph{Ranker} puts each activity into several different queues according to the IP address of its \emph{context identifier}. Naturally, activities in the same queue are sorted according to the same local clock, so \emph{Ranker} only need to compare head activities of each queue and select candidate activities for composing CAGs according to the following rules:

\textbf{Rule 1}: If a head activity $A$ in a queue has RECEIVE type and \emph{ranker} had found an activity $X$ in the \emph{mmap}, of which $X \rightarrow_{m} A$ holds true, then $A$ is the candidate.

If a key is \emph{the message identifier} of an activity $A$ and the value of the \emph{mmap} points to a SEND activity $X$ with the same message identifier, we can say $X \rightarrow_{m} A$.

\textbf{Rule 1} ensures that when a SEND activity has became a candidate and been delivered to \emph{engine}, the RECEIVE activity having message relation with it will also become a candidate once it becomes a head activity in its queue.

\textbf{Rule 2}: If no head activity is qualified with \textbf{Rule 1}, then \emph{ranker} compares the type of head activities in each queue according to the priority of $BEGIN \prec SEND \prec END \prec RECEIVE \prec MAX$. Finally the head activity with the lower priority is the candidate.

\textbf{Rule 2} ensures that a SEND activity $X$ always becomes a candidate earlier than a RECEIVE activity $A$, if $X \rightarrow_{m} A$ holds true.

After a candidate activity is chosen, it will be popped out from its queue and delivered to \emph{engine}, and \emph{engine} matches the candidate with an incomplete CAG. Then the element next to the popped candidate will become a new head activity in that queue. At the same time, \emph{ranker} will update the new minimal timestamp in the sliding time window and fetch new qualified activities into the buffer of \emph{ranker} in a new round.

\subsubsection{Constructing CAG} \label{section_construct_cags}
\emph{Engine} fetches a candidate, outputted by \emph{Ranker}, and matches it with an incomplete CAG. The following pseudo-code illustrates the correlation algorithm. In line 1, engine iteratively fetches a candidate activity \emph{current} by calling \emph{rank() function} of \emph{Ranker}, introduced in Section \ref{section_choose_activity}. From line 2-34, \emph{Engine} parses and handles activity \emph{current} according to its activity type. Line 3-11 handles \emph{BEGIN and END activities}. For \emph{BEGIN activity}, a new CAG is created. For \emph{END activity}, the construction of its matched CAG is finished.

\textbf{Procedure} correlate \textbf{\{}\\
1:~\textbf{while} (\emph{current}=ranker.rank ( )) {\\
2:~~~\textbf{switch} (current$\rightarrow$get\_type ( )) {\\
3:~~~~~\textbf{case} BEGIN:\\
4:~~~~~~~create a CAG with \emph{current} as its root;\\
5:~~~~~\textbf{case} END:\\
6:~~~~~~~find the matched parent where parent$\rightarrow_{c}$current;\\
7:~~~~~~~\textbf{if} (the match is found) {\\
8:~~~~~~~~~add \emph{current} into the matched CAG;\\
9:~~~~~~~~~add an \emph{adjacent context edge} from \emph{parent} to \emph{current};\\
10:~~~~~~~~output CAG;\\
11:~~~~~\}\\
12:~~~~\textbf{case} SEND:\\
13:~~~~~~find matched parent\_msg where parent\_msg$\rightarrow_{c}$current;\\
14:~~~~~~\textbf{if} (the match is found) \{\\
15:~~~~~~~~\textbf{If} ( parent\_msg.type==SEND£©\textbf{\}}\{\\
16:~~~~~~~~~~parent\_msg.size $+=$ current.size;\\
17:~~~~~~\textbf{else}\\
18:~~~~~~~~~~add \emph{current} into the matched CAG.\\
19:~~~~~~~~~~add an \emph{adjacent context edge} from parent\_msg to \emph{current}.\\
20:~~~~~~~~\textbf{\}}\\
21:~~~~~~\}\\
22:~~~~\textbf{case} RECEIVE:\\
23:~~~~~~find matched parent\_msg where parent\_msg$\rightarrow_{m}$current;\\
24:~~~~~~\textbf{if} (the match is found) \{\\
25:~~~~~~~~parent\_msg.size-=current.size;\\
26:~~~~~~~~\textbf{if} (parent\_msg.size $==$0) \{\\
27:~~~~~~~~~~add \emph{current} into the matched CAG;\\
28:~~~~~~~~~~add a \emph{message edge} from parent\_msg to \emph{current};\\
29:~~~~~~~~~~find matched parent\_cntx where parent\_cntx$\rightarrow_{c}$current;\\
30:~~~~~~~~~~\textbf{if} (the match is found)\\ 	
31:~~~~~~~~~~~~\textbf{if} (parent\_msg and parent\_cntx are in the same CAG)\\
32:~~~~~~~~~~~~~~add a \emph{context edge} from parent\_cntx to \emph{current};\\
33:~~~~~~~~\}\\
34:~~~~~~\}\\
35:~~\}//switch\\
36:\}//while\\
37:\textbf{\}}//correlate\\

Line 12-34 handle SEND and RECEIVE activities. Activities are inherently asymmetric between a sender and a receiver because of their underlying buffer sizes and delivery mechanism, and hence a match between \emph{SEND} and \emph{RECEIVE} activities is not always one-to-one, but \emph{n-to-n} relations. Fig.\ref{fig_n-n_message} shows a case that a sender consecutively sends a message in two parts and a receiver receives messages in three parts. Our algorithm correlates and merges these activities according to \emph{message sizes} in \emph{message identifiers}.
\begin{figure}[hbtp]
  \centering
  \includegraphics[scale=0.45]{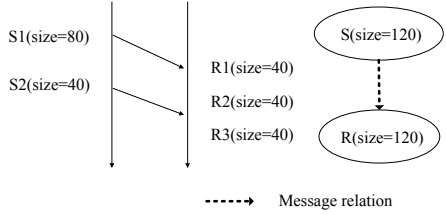}\\
  \caption{Merging multiple SEND and RECEIVE activities}
  \label{fig_n-n_message}
\end{figure}
A situation may happen that an activity is wrongly correlated into two causal paths because of reusing threads in some concurrent programming paradigms. For example in a thread-pool implementation, one thread may serve one request at a time. When the work is done, the thread is recycled into the thread pool. Line 30-31 check if the two parents are in the same CAG. If the check returns true, \emph{Engine} will add an edge of \emph{context relation}, or else not.

\subsubsection{Disturbance tolerance}\label{section_disturbance_tolerance}
In a clean environment without disturbance, our algorithm in Section \ref{section_choose_activity} and Section \ref{section_construct_cags} can produce correct causal paths. But in a practical environment, there are many disturbances. In the rest of this subsection, we consider how to resolve \emph{noise activities disturbance}, \emph{concurrency disturbance}, and \emph{clock skew disturbance}.

\textbf{Noise activities disturbance}: \emph{Noise activities} are caused by other applications coexisting with the target service on the same nodes. Their activities through the kernel's TCP stack will also be logged and gathered by our tool.

\emph{Ranker} handles noise activities in two ways: 1) filters noise activities according to their attributes, including program name, IP and port. 2) If activities can not be filtered with the attributes, the ranker checks them with \emph{is\_noise()} function. If true, the ranker will discard them. The \emph{is\_noise()} function is illustrated in the following pseudo codes.\\
bool \textbf{is\_noise} (Activity * E) \{\\
return (( E$\rightarrow$type==RECEIVE)\&\&(No matched SEND activity X in \emph{mmap} with X$\rightarrow_{m}$E) \&\& ( No matched SEND activity Y in the buffer of ranker with Y$\rightarrow_{m}$E));\\
\}

\textbf{Concurrency disturbance}: The second disturbance is called \emph{concurrency disturbance}, which only exists in multi-processor nodes. Fig. \ref{fig_8}-a illustrates a possible case, of which two concurrent requests are concurrently served  by two multi-processor nodes and four activities are observed. $S^{1,1}_{1,2}$ means a SEND activity produced on the \emph{CPU1} of \emph{Node1}, and $R^{2,0}_{1,2}$ is its matched RECEIVE activity produced on the \emph{CPU0} of \emph{Node2}. When these four activities are fetched into the buffer of \emph{Ranker}, they are put into two queues as shown in Figure. \ref{fig_8}a. The head activities of both two queues are RECEIVE activities and hence block the matched SEND activities of each other. \emph{Ranker} handles this case by swapping the head activity and its following activity in the first queue. Figure. \ref{fig_8}b illustrates our solution.
\begin{figure}[hbtp]
  \centering
  \includegraphics[scale=0.50]{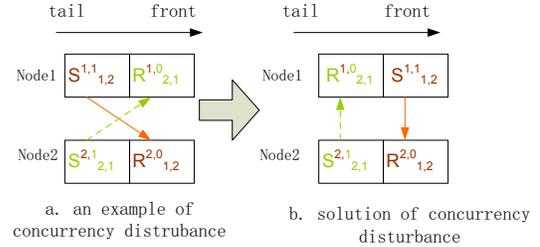}\\
  \caption{Example of concurrency disturbance}
  \label{fig_8}
\end{figure}

\textbf{Clock skew disturbance}: As explained in Section \ref{section_choose_activity}, activities will be fetched into the buffer of \emph{Ranker} according to their local timestamp. \emph{RECEIVE activities may be fetched into the buffer before their corresponding SEND activities when local clock of SEND activities is much later than RECEIVE activities because of clock skew}. We take a simple solution to resolve this issue. In comparing head activities of each queue, we record the timestamp of the first activity from each node, and then we can calculate the approximate clock skew between two nodes. According to the the approximate clock skew, we remedy timestamp of activities on the node with the larger clock skew, and hence we can prevent the case mentioned above from happening.

\subsection{Improving systems scalability} \label{section_scalability}
In this section, we present two mechanisms to improve system scalability.

\subsubsection{Tracing on demand}

The instrumentation mechanism of \emph{PreciseTracer} depends on a open source software named \emph{SystemTap} \url{http://sourceware.org/systemtap}, which extends the capabilities of \emph{Kprobe} \cite{Kprobe}- a tracing tool on a single Linux node. Using \emph{SystemTap}, we have written the \emph{LOG\_TRACE} module, which is a part of \emph{Gather}. \emph{LOG\_TRACE} could obtain context information of processes and threads from the operating system and inserts probe points into \emph{tcp\_sendmsg()} and \emph{tcp\_recvmsg()} functions of the kernel communication stack to log sending or receiving activities.

Deployed on every node, \emph{Gather} receives commands from \emph{Coordinator}. When users demands \emph{PreciseTracer} to be enabled or disabled on demand, \emph{Coordinator} will synchronize each \emph{Gather} to dynamically load or unload the kernel module \emph{LOG\_TRACE}, which are supported by the Linux OS. \emph{PreciseTracer} could choose instrumentation mode of \emph{continuous collection} or \emph{Tracing on demand} or \emph{periodical sampling}. When administrators find services are abnormal, they can choose the model of \emph{Tracing on demand}, which starts tracing requests according to commands of administrators. When administrators have pinpointed the problems, they can stop tracing requests. When \emph{PreciseTracer} choose the mode of \emph{periodical sampling}, it will be enabled or disabled alternatively, which could decrease the overhead on running applications, and hence improve system scalability.

\subsubsection{sampling} \label{section_sampling}
To consider large amount of logs produced by tracing requests of multi-tier services, it seems that \emph{sampling} is a straightforward solution. However, it is not a trivial issue to support sampling in accurate requesting tracing approaches: first, the tracing mechanism must allow to be enabled or disabled on demand; second the tracing algorithm must tolerate loss of logs. In this section , we discuss how to tolerate losses of activities and consider three cases:

\begin{itemize}
  \item \textbf{Case 1}: Lost BEGIN and END activities;
  \item \textbf{Case 2}: Lost RECEIVE activities;
  \item \textbf{Case 3}: Lost SEND activities.
\end{itemize}

It is difficult to handle Case 1. Each CAG needs a BEGIN activity and an END activity to identify its begin and end. Fortunately, losses of BEGIN and END activities only affect constructing their affiliated CAG, and have no influence on other CAGs that have identified BEGIN and END activities.

About Case 2, due to the underlying delivery mechanism, a receiver will receive a message in several parts, which is explained in Section \ref{section_construct_cags}. So the situation seldom happens that all parts of a message fail to be collected. In this case, the \emph{received message size} will be less than its corresponding \emph{sent message size}, but this wouldn't prevent from constructing a CAG.

\begin{figure}[hbtp]
  \centering
  \includegraphics[scale=0.44]{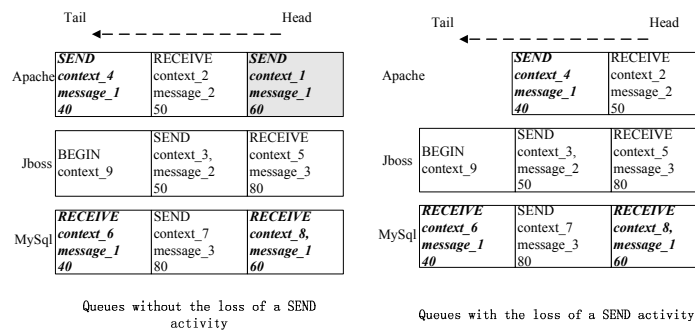}\\
  \caption{A case of lost SEND activities}
  \label{fig_lost_of_logs}
\end{figure}

Fig. \ref{fig_lost_of_logs} shows a case of lost SEND activities when candidate activities are in queues of  \emph{Ranker}. Activities from the same node are put into a queue and ordered according to their local timestamp. We hereby utilize \emph{(activity type, context identifier, message identifier, message size}) to identify an activity. In Fig. \ref{fig_lost_of_logs}, \emph{(SEND, context\_1, message\_1, 60)} is related to \emph{(RECEIVE, context\_8, message\_1, 60)} while \emph{(SEND, context\_4, message\_1, 40)} is related to \emph{(RECEIVE, context\_6, message\_1, 40)} and they share the same \emph{message identifier}. \emph{Ranker} would pick candidate activities. However, if it fails to collect activity \emph{(SEND, context\_1, message\_1, 60)}, activity types of all head activities will be RECEIVE, and our algorithm mentioned above can not proceed. We propose a simple approach to resolve this issue, and we just discard the RECEIVE activity with the smallest timestamp. In Section \ref{section_evaluating_accuracy}, our experiments show our approach to handling \emph{lost activities} is acceptable.


\subsubsection{The complexity of the algorithm} \label{section_algorithm_complexity}
For a multi-tier service, the time complexity of our algorithm is approximately $O(g*p*\Delta n)$, where $g$ measures the structure complexity of a service, $p$ is the number of requests in the fixed duration, $\Delta n$ is the size of activities sequence per request in the sliding time window. Furthermore, the time complexity of our algorithm can be expressed as $O(g*n)$, where n is the size of activities sequence in the sliding time window. The space complexity of our algorithm is approximately $O(2g*p*\Delta n)$ or $O(2g*n)$.

\subsection{PreciseTracer Implementation} \label{implementation}

We have implemented \emph{PreciseTracer} with the following components: \emph{TCP\_Tracer, Correlator, Analyzer} and a visualized tool named \emph{PathViewer}.

If the kernel module named \emph{LOG\_TRACE}  (a part of \emph{Gather}) is loaded, when an application sends or receives a message, a probe point will be trapped and an activity is logged.
The original format of an interaction activity produced by \emph{LOG\_TRACE} is \emph{"timestamp hostname program\_name Process ID Thread ID SEND/RECEIVE sender\_ip: port-receiver\_ip: port message\_size"}. \emph{Gather} transforms the original format of an interaction activity into more understandable n-ary tuples to describe \emph{context and message identifiers} of activities, described in Section \ref{TCP_Tracer}. Distinguishing activity types is straightforward. SEND and RECEIVE activities are transformed directly. BEGIN or END activities are distinguished according to the port of the communication channel. For example, the RECEIVE activity from a client to the web server's port 80 means the START of a request, and the SEND activity in the same connection with opposite direction means the STOP of a request. After all \emph{Gathers} have finished transformation, \emph{Coordinator} would start \emph{Correlator} and \emph{Analyzer} to deal with collected logs.

\emph{Correlator} constructs CAGs and delivers them to \emph{Analyzer}. \emph{Analyzer} analyzes CAGs to obtain causal path patterns, and further calculates statistical information of those patterns.

Unlike \emph{Analyzer}, which only provides administrators with analysis results, \emph{PathViewer} is a graphical user interface, which provides administrators with visual views of causal paths. Presently, we provides two view of CAGs: (1) time-space diagram of CAGs; and (2) causal path pattern of CAGs. Fig. \ref{fig_pathviewer} shows a picture of the time-space diagram of a CAG, from which administrator would get detail information about how this request was served by multi-tier services.

\begin{figure}[hbtp]
  \centering
  \includegraphics[scale=0.45]{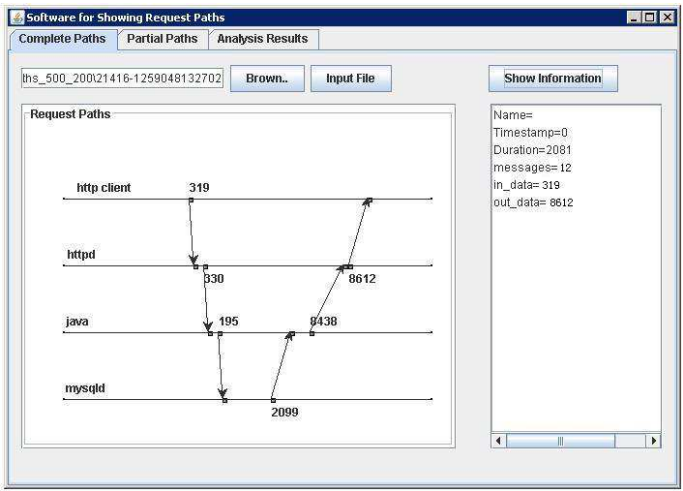}\\
  \caption{The time-space diagram of serving a request}
  \label{fig_pathviewer}
\end{figure}

\section{Evaluation} \label{evaluation}
In this section, we will evaluate \emph{PrecisTracer}. First, we introduce the experimental environment and setup; second, we evaluate the accuracy of our tracing tool; third, we test the efficiency of the tracing algorithm; fourth, we demonstrate the online analysis ability of \emph{PreciseTracer}; fifth, we present the sampling effect; finally, we demonstrate how \emph{PreciseTracer} facilitates debugging performance.

\subsection{Experimental environment and setup}
We have performed experiments on RUBiS and TPC-W\cite{TPCW}. In the rest of experiments, We choose RUBiS as the target application. Developed by Rice University, RUBiS is a three-tier auction site prototype modeled after eBay.com, which is used to evaluate application servers' performance scalability.

\begin{figure}[hbtp]
  \centering
  \includegraphics[scale=0.40]{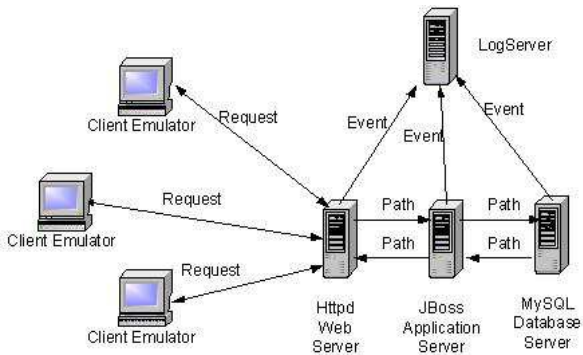}\\
  \caption{The deployment diagram of RUBiS}
  \label{fig_deployment}
\end{figure}

The experiment platform is a 6-node Linux cluster connected by a 100Mbps Ethernet switch. Web tier (Apache) and active web pages server (JBOSS) is respectively deployed on two SMP nodes with two PIII processors and $2 G$ memory. Database (MYSQL) and the analysis component of \emph{PreciseTracer} are respectively deployed on two SMP nodes with eight Intel Xeon processors and $8 G$ memory. Every node runs the Redhat Federo Core 6 Linux with the kprobe \cite{Kprobe} feature enabled. The deployment of RUBiS is shown in Fig. \ref{fig_deployment}.

In the following experiments, client nodes emulate two kinds of workload: read\_only workload (Brows\_only) and read\_write mixed workload (Default). We utilize two nodes to emulate clients. Each node is set to the same number of concurrent clients. In order to make the experiences close to reality, we adapt mix workload in all experiences.

According to the user guide of RUBiS, every workload includes 3 stages: up ramp, runtime session, and down ramp. In the following experiences, we set different durations for three stages.

\subsection{Evaluating \emph{PreciseTracer} accuracy} \label{eva_accuracy}

\label{section_evaluating_accuracy}
To verify the accuracy of \emph{PreciseTracer}, we build another library-interposition tool to collect network communication. We rewrite the following system library functions in our library: \emph{write, writev, send, read, recv}. When a message is sent, a global request ID will be tagged and propagated, and hence we can obtain causal paths with 100\% accuracy. The following attributes are logged for the Apache web server, the JBoss Server and the MySql database, including (1) request ID, (2) the start time and end time of serving a request; (3) ID of the process or thread.

At the same time, with only application independent knowledge, such as \emph{timestamps}, \emph{end-to-end
communication channels}, we use \emph{PreciseTracer} to obtain causal paths. For each causal path, we
independently obtain information like (2) and (3). If all attributes of a causal path are consistent with the ones obtained from our another library-interposition tools, we confirm that the causal path is correct. So we define the path accuracy as follows:

\emph{Path accuracy = correct paths/ all logged requests}
We test the accuracy of our algorithm in the \emph{offline} mode with the configuration of \emph{(offline, 20 milliseconds)} for read\_write mixed workload of RUBiS. The configuration \emph{(offline, 20 milliseconds)} indicates the sliding time windows is 20 milliseconds. The number of concurrent clients is respectively set as 200, 500 and 800. The up ramp duration, runtime session and down ramp duration is respectively set as  1 minute, 5 minutes and 1 minute. Table \ref{table_2} summarizes the experiment reports, and the accuracy of our algorithm are almost 100\%. This is because most of activities of components are logged using \emph{SystemTap} (which fails to collect a negligible fraction of logs), so our algorithm almost correlates all activities into causal paths.






\begin{table}[hbtp]
\renewcommand{\arraystretch}{1.3}
\caption{CAGs' Accuracy Results}
\label{table_2}
\centering
\begin{tabular}{|l||l|l|l|}
  \hline
  & \itshape 200 & \itshape 500 & \itshape 800 \\ \hline\hline
  \itshape Total CAGs (library interposition) & 3617 & 13627 & 10403 \\ \hline
  \itshape matched & 3610 & 13623 & 10402 \\ \hline
  \itshape accuracy & 99.97\% & 99.81\% & 99.99\% \\
  \hline
\end{tabular}
\end{table}


Since we adopt a sampling policy to increase system scalability, we can not accurately keep complete logs, which constitutes different causal paths, so we delete logs randomly to test the ability of our algorithm to handle lost logs. We conduct four groups of experiments: only deleting logs from \emph{Apach}, only deleting logs from \emph{Jboss}, only deleting logs from \emph{MySql}, and deleting logs from all three components. The deletion of logs is conducted randomly according to a defined percent. The number of concurrent clients is 800. We adapt read\_write mixed workload and each workload is set as followed: up ramp with a duration of 2 minutes, runtime session with a duration of 5 minutes, and down ramp with a duration of 1 minute.
\begin{figure}[hbtp]
  \centering
  \includegraphics[scale=0.45]{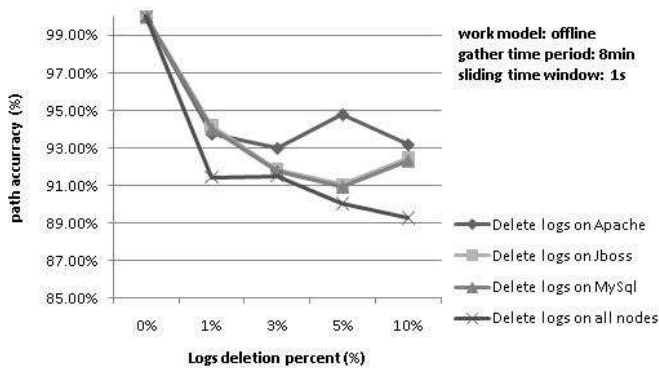}\\
  \caption{Path accuracy v.s. deletion percent}
  \label{fig_robust_1}
\end{figure}


Fig. \ref{fig_robust_1} shows the deletion percent's impact on the path accuracy. Because many tuple logs are deleted, some CAGs couldn't be produced rightly. From Fig. \ref{fig_robust_1}, even under the worst situation (deleting logs from all component) and the deletion percent is as high as 10\%, the path accuracy is about 90\%.


\subsection{Evaluating \emph{PreciseTracer} efficiency}
In this section, we respectively evaluate the complexity and the overhead of \emph{PreciseTracer}. Lastly, we demonstrate its online analysis ability.


\subsubsection{Evaluating the complexity} \label{eva_complex}


We set the baseline configuration as \emph{(online, 120, 400, 20, 1)} in this experiment, and the detailed parameter analysis will be deferred to the end of this subsection. This configuration \emph{(online, 120, 400, 20, 1)} indicates that \emph{PreciseTracer} will work under online model; It will gather logs for one round, lasting for 120 seconds (collecting time window) and then stop collecting logs (in next 400 seconds); The sliding time window is 20 milliseconds.

When concurrent clients vary from 100 to 1000, we record the number of served requests and the correlation time. For different numbers of concurrent clients, the test duration is fixed for the read\_write mixed workload. And each workload is set as following: up ramp with the duration of 2 minutes, runtime session with the duration of 5 minutes, and down ramp with the duration of 1 minute.
\begin{figure}[hbtp]
  \centering
  \includegraphics[scale=0.44]{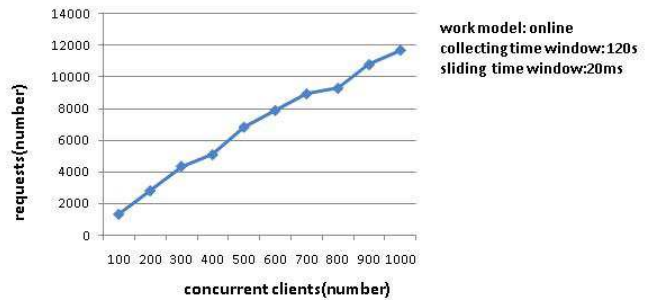}\\
  \caption{Requests v.s. concurrent clients}
  \label{fig_12}
\end{figure}

\begin{figure}[hbtp]
  \centering
  \includegraphics[scale=0.44]{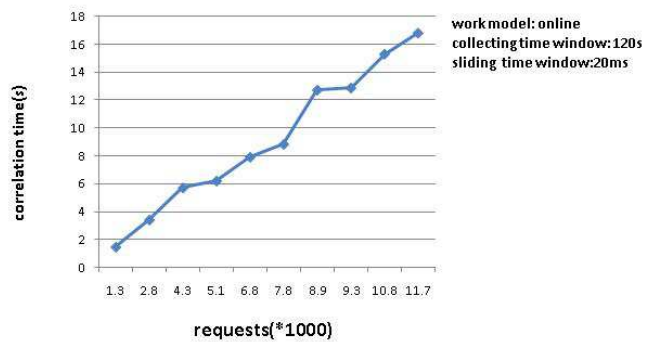}\\
  \caption{Correlation time v.s. requests (the unit in x-axis is 1000 requests) }
  \label{fig_13}
\end{figure}

From Fig. \ref{fig_12} and Fig. \ref{fig_13}, we observe that \emph{the number of requests is almost linear in the number of concurrent clients} and \emph{the correlation time is almost linear in the number of requests in the fixed duration}. In Section \ref{section_algorithm_complexity}, we conclude that the time complexity of our algorithm is approximately $O(g*p*\Delta n)$. Our experiment data in Fig. \ref{fig_13} is consistent with this analysis, since $g$ is a constant for RUBiS and $\Delta n$ is unchanged in the fixed sliding time window, so the correlation time is linear in the number of requests in the fixed test.

There are two important parameters in experiments, which could influence the efficiency of our correlate algorithm: \emph{collecting time window} and \emph{sliding time window}. A \emph{collecting time window} is the time duration of gathering logs. So the longer collecting time window is, the more logs that our algorithm should deal with at a time and the higher time cost for our system. The influence of \emph{the sliding time window} is much more complex, and Fig. \ref{fig_14} shows the effect of different size of the sliding time window on the correlation time for different numbers of concurrent clients (200, 500, and 800).
\begin{figure}[hbtp]
  \centering
  \includegraphics[scale=0.45]{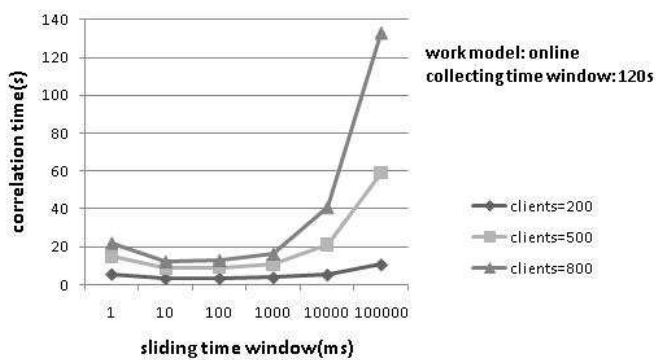}\\
  \caption{Correlation time v.s. sliding time window}
  \label{fig_14}
\end{figure}

The size of sliding time window would effect the correlation time in two ways: first, when the number of requests in the fixed duration is unchanged, the time complexity of the algorithm is linear in the size of $\Delta n$ for RUBiS. $\Delta n$ is \emph{the size of activities sequence per request in the sliding time window}, which is determined by the size of the sliding time window. So the correlation time will increase with the sliding time window; secondly, the sliding time window also affects handling noise activities. If the size of the sliding time window is set to be a smaller value  (less than 10 milliseconds), it has to deal with noise activities  more frequently, and hence increases the correlation time. From Fig. \ref{fig_14}, we can observe those effects.
\begin{figure}[hbtp]
  \centering
  \includegraphics[scale=0.45]{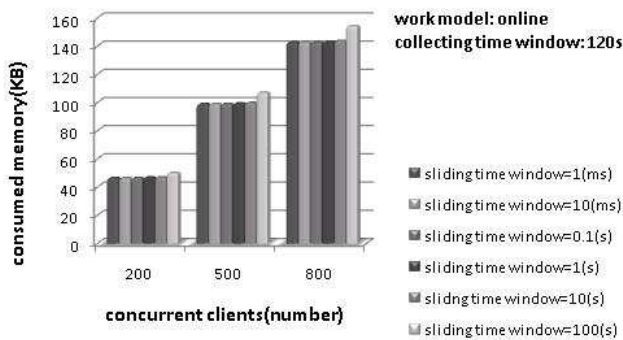}\\
  \caption{Memory consumption v.s. sliding time window}
  \label{fig_15}
\end{figure}

Fig. \ref{fig_15} shows the effect of the size of the sliding time window on the memory consumption of \emph{Correlator} for different concurrent clients. We do not show the memory consumption of other components, because they consume fewer memory in comparison with that of \emph{Correlator}.

\subsubsection{The overhead of \emph{PreciseTracer}} \label{eva_overhead}
We compare the throughput and average response time of RUBiS for the read\_write mixed workload when the instrumentation mechanism is disabled (no instrumentation), enabled (only starting \emph{Gather} module) or \emph{PreciseTracer} is in the model of \emph{online analysis}. In order to test the maximum overhead of PreciseTracer under online model, we choose the \emph{continuous collection} mode. Thus, we set the configuration of \emph{PreciseTracer} as \emph{(online, 60, 0, 20, 7)}. Every workload is set as following: up ramp with the duration of 2 minutes, runtime session with the duration of 5 minutes, and down ramp with the duration of 1 minute.
\begin{figure}[hbtp]
  \centering
  \includegraphics[scale=0.45]{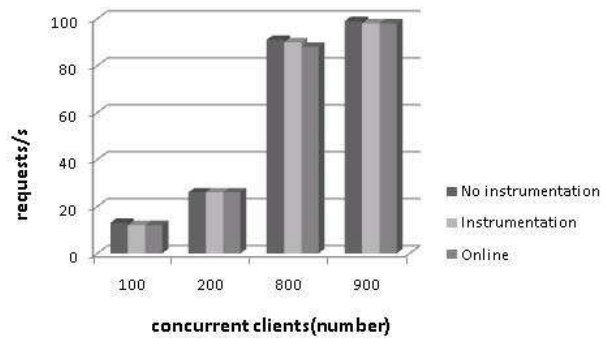}\\
  \caption{The effect on the throughput in terms of requests/s}
  \label{fig_20}
\end{figure}
\begin{figure}[hbtp]
  \centering
  \includegraphics[scale=0.45]{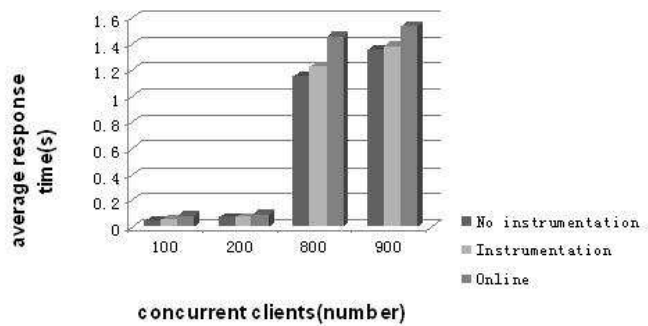}\\
  \caption{The effect on the average response time}
  \label{fig_21}
\end{figure}

In Fig. \ref{fig_20} and Fig. \ref{fig_21}, we observe that \emph{PreciseTracer} in the mode of \emph{online analysis} have little effect on the throughput and small effect on the average response time of RUBiS.

\subsection{Evaluating the online analysis ability of PreciseTracer}
In this section, we demonstrate the online analysis ability of \emph{PreciseTracer}. We set the baseline configuration as \emph{(online, 60, 0, 20, 10)}. The test duration is fixed for the read\_write mixed workload. The workload is set as followed: up ramp with the duration of 2 minutes, runtime session with the duration of 5 minutes, and down ramp with the duration of 1 minute.
\begin{figure}[hbtp]
  \centering
  \includegraphics[scale=0.45]{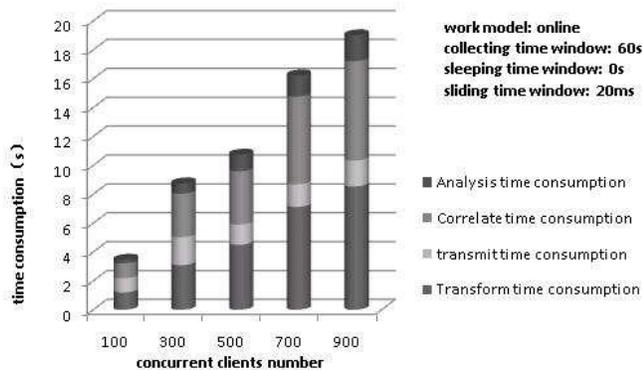}\\
  \caption{Time consumption in each stage}
  \label{fig_24}
\end{figure}


There are four steps in applying request tracing for online analysis of services: first, the system would transform original logs into tuple logs; secondly, it will transmit tuple logs to \emph{a analysis node} for further work; then the system correlates those tuple logs into CAGs; finally, the tool analyzes useful information from those CAGs. So the time consumption of PreciseTracer include four main stages: transform time, transmit time, correlate time and analysis time. Because the second step is conducted on each node simultaneously, we take the maximum one as the time consumption in this stage.

Fig. \ref{fig_24} shows that \emph{PreciseTracer} could deal with logs efficiently. It can output analysis results within 20 seconds even the workload is so heavy as 900 concurrent clients. This demonstrates that our system could be applied in online analysis.

\subsection{The sampling effect}
\emph{PreciseTracer} supports sampling through tolerating losses of activities. In this section, through experiments, we demonstrate that adopting a sampling policy could decrease the size of logs to be collected and analyzed, at the same time we still can capture most of dominated causal path patterns.

We run experiments twice with the baseline configurations \emph{(offline, 20 milliseconds)} respectively for one minutes and 10 minutes. The test duration is fixed for the brown\_only workload. And every workload is set as following: up ramp with the duration of 30 seconds, runtime session with the duration of 30 minutes, and down ramp with the duration of 1 minute.

We analyze top 10 dominated causal path pattern in two run of experiments (1 minute V.S. 10 minutes). We have two observations: first, in two runs, causal paths belonging to the top 10 dominated causal patterns respectively take up a significant percent of all causal paths (both 88\%); Second, two runs of experiments have the same top 10 dominated causal path patterns. These two observations justified adopting the sampling policy, since the adoption of the sampling policy still can capture most of dominated casual path patterns, even not all causal path patterns.

\begin{figure}[hbtp]
  \centering
  \includegraphics[scale=0.46]{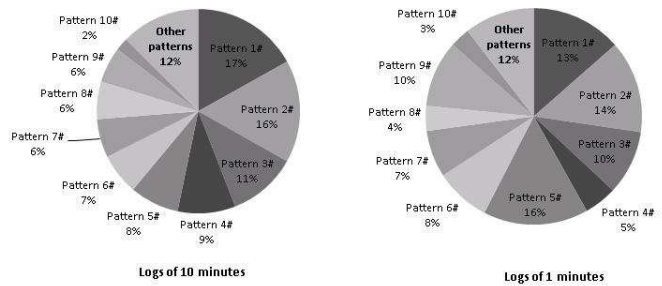}\\
  \caption{Comparisons of dominated causal path patterns in two runs respectively of one minute and 10 minutes logs}
  \label{fig_sampling}
\end{figure}

Table \ref{table_sampling} compares log sizes in two runs of experiments respective with 1 minute and 10 minutes. It indicates that sampling captures most of dominated causal path patterns, at the same time  decreases the cost of collecting and analyzing logs, and hence improves system scalability.
\begin{table}[hbtp]
\renewcommand{\arraystretch}{1.3}
\caption{The log sizes in two runs of experiments with 1 minute and 10 minutes}
\label{table_sampling}
\centering
\begin{tabular}{|l|p{1.5cm}|p{1.5cm}|p{1.5cm}|}
  \hline
  & \itshape Original logs on Apache (M) & \itshape Original logs on Jboss (M) & \itshape Original logs on MySql (M) \\ \hline
  \itshape 1 minute & 3.9 & 6.6 & 6.7 \\ \hline
  \itshape 10 minutes & 27.9 & 55.1 & 58.6 \\ \hline
  \hline
\end{tabular}
\end{table}

\subsection{Identifying performance bottleneck}
In the following section, we will utilize \emph{PreciseTracer} to detect performance problems of multi-tier service. The following experiments are done offline. The experiment platform is a Linux cluster connected by a 100Mbps Ethernet switch. Web tier (APACHE) and active web pages server (JBOSS), database (MYSQL) and analysis components of \emph{PreciseTracer} is respectively deployed on four SMP nodes, each of which  has two PIII processors and $2 G$ memory. Each node runs the Redhat Federo Core 6 Linux with the kprobe \cite{Kprobe} feature enabled. The deployment of \emph{RUBiS} is similar to Fig. \ref{fig_deployment}.

\subsubsection{Misconfiguration shooting}

When we perform offline experiments using \emph{PreciseTracer} with the configuration of \emph{(offline, 20 milliseconds)}, we observe that when the number of concurrent clients increases from 700 to 800, the throughput of RUBiS decreases. Fig. \ref{fig_26} shows the relationship between number of concurrent clients and requests The test duration is fixed for the brown\_only workload. And every workload is set as following: up ramp with the duration of 1 minutes, runtime session with the duration of 7 minutes, and down ramp with the duration of 1 minute. An interesting question is what is the wrong with RUBiS?

\begin{figure}[hbtp]
  \centering
  \includegraphics[scale=0.45]{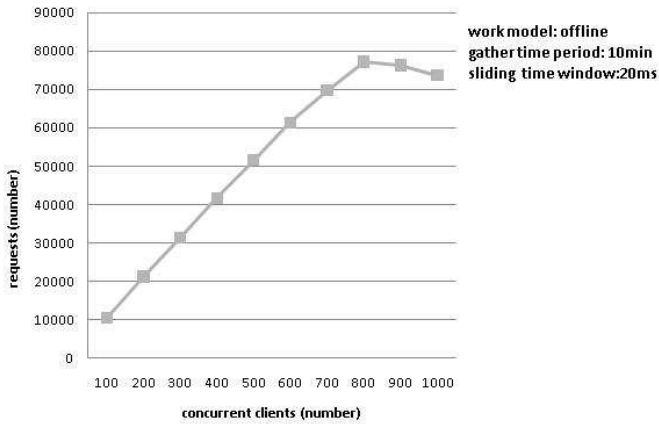}\\
  \caption{Requests v.s. concurrent clients}
  \label{fig_26}
\end{figure}

Generally, we will observe the resource utilization rate of each tier and the metrics of quality of service to pinpoint the bottlenecks. Using the monitoring tool of RUBiS, we notice that the CPU usage of the each node is less than 80\% and the I/O usage rate is not high. Obviously, the traditional method does not help.

To answer this question, we use our tool to analyze the most frequent request ViewItem for RUBiS, compute the most dominated causal path pattern and visualize the view of \emph{latency percentages of components}. We identify the problem quickly.
\begin{figure}[hbtp]
  \centering
  \includegraphics[scale=0.45]{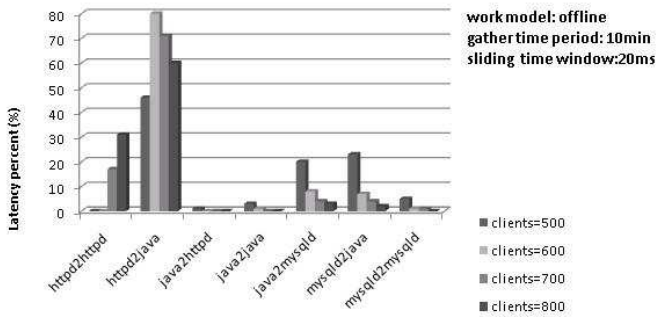}\\
  \caption{The latency percentage of components}
  \label{fig_27}
\end{figure}

From Fig. \ref{fig_27}, we observe that when the number of concurrent clients increases from 500 to more, the latency percentage of httpd2Java from the first tier to the second tier changes dramatically, and the value is 46\%, 80\%, 71\% and 60\% respectively for 500, 600, 700 and 800 concurrent clients. In Fig.\ref{fig_27}, the latency percentage of httpd2Java is 46\% for 500 clients, which means that the processing time of the interaction from httpd to Java takes up 46\% of the whole time of servicing a request.

At the same time, the latency percentage of httpd2httpd (first tier) increases dramatically from 17\% (700 clients) to 31\% (800 clients). We observe the CPU usage of the Jboss node is less than 60\% and the I/O usage rate is not high. When servicing a request, the httpd2httpd is before the httpd2java in a causal path. So we can confirm that there is something wrong with the interaction between the httpd and the JBoss. Through reading the manual reference of RUBiS, we confirm that the problem may be mostly related with the configuration of thread pool in the JBoss. According to the manual book of the JBoss, one parameter named MaxThreads controls the max available thread number, and one thread services a connection. The default value of MaxThreads is 40.

We set MaxThreads as 250 and run the experiments again. In Fig. \ref{fig_28}, we observe that our work is effective. When concurrent clients increase from 500 to 800, the throughput is increased and the average response time is decreased in comparison with that of the default configuration. However, for 900 concurrent clients, the resource limit of hardware platform results in the new bottleneck. In Fig. \ref{fig_28}, TP\_MT40 is the throughput when MaxThreads is 40, and RT\_MT250 is the average response time when MaxThreads is 250.
\begin{figure}[hbtp]
  \centering
  \includegraphics[scale=0.45]{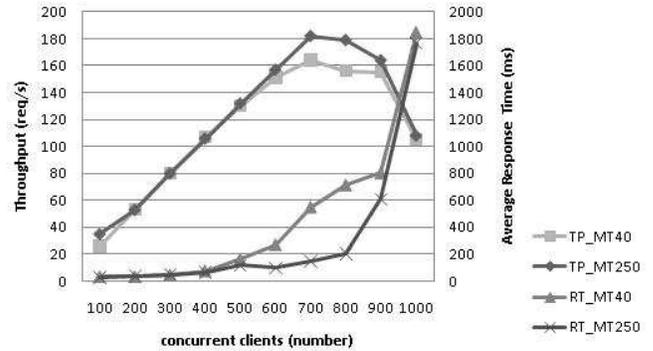}\\
  \caption{Performance for different MaxThreads}
  \label{fig_28}
\end{figure}

\subsubsection{Injected performance problem} \label{injected_problems}
To further validate the accuracy of locating performance problems using PreciseTracer, we have injected several performance problems into RUBiS components and the host nodes: for abnormal case 1, we modify the RUBiS code to inject a random delay into the second tier; for abnormal case 2, we lock the items table of the RUBiS database to inject a delay into the third layer; for abnormal case 3, we change the configuration of the Ethernet driver from 100 M to 10 M on the node running the JBoss.
\begin{figure}[hbtp]
  \centering
  \includegraphics[scale=0.45]{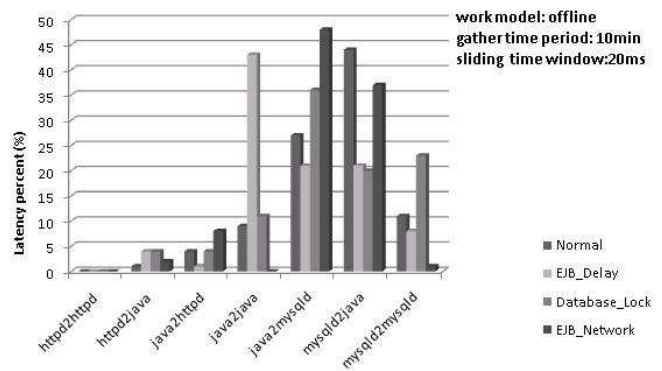}\\
  \caption{Latency percentages of components for abnormal cases}
  \label{fig_29}
\end{figure}

We use PreciseTracer to locate the component in question where different performance problems are injected. Fig. \ref{fig_29} shows the latency percentages of components for normal case and three abnormal cases.

For abnormal case 1 (EJB\_Delay), the latency percent-age of Java2Java (JBoss, second tier) increases from less than 10\% for the normal case to more than 40\%, and the latency percentages of other components decrease with different amounts. So we can confirm that JBoss is in question.

For abnormal case 2 (DataBase\_Lock), the latency percentage of mysqld2mysqld (third tier) increases from 12\% for the normal case to more than 20\%, and the latency percentage of java2mysqld (interaction from second tier to third tier) increases from 26\% for the normal case to more than 35\%. The Latency percentages of other components keep unchanged or decrease. So we confirm that MySQL is in question.

For the abnormal case 3 (EJB\_Network), the latency percentage of Java2mysqld (from second tier to third tier) increases from 26\% for the normal case to 47\%; mysqld2java (from third tier to second tier) keeps about 37\%. The latency percentage of httpd2java from first tier to second tier increases from 1\% to 2\%; the percentage of java2httpd from second tier to first tier increases from 4\% to 8\%. We observe that most of time of servicing request is spent on the interactions between second tier and third tier, and the three latency percentages of four interactions related with the second tier are increased. We confirm the second tier is in question. Further observation shows the latency percentage of Java2java strangely decreases from 9\% to almost 0\%. So we confirm that there is something wrong with the network of second tier.

\section{Conclusion} \label{conclusion}
We have developed an accurate request tracing tool: \emph{PreciseTracer} to help users understand and debug performance problems of a multi-tier service of black boxes. Our contributions are two-fold: (1) we have designed a precise tracing algorithm to derive causal paths for each individual request, which only uses the application-independent knowledge, such as timestamps and end-to-end communication channels; (2) We respectively presented abstractions: \emph{component activity graphs} and \emph{dominated causal path patterns} for understanding and debugging micro-level and macro-level user request behaviors of services; (3) we have presented two mechanisms: \emph{tracing on demand} and \emph{sampling} to increase system scalability, and our experiments show sampling efficiently preserves performance data of services in the way that it captures most of \emph{dominated causal path patterns}, at the same decreases the cost of collecting and analyzing logs; (4) we have designed and implemented an online request tracing system. Our experiments demonstrated that \emph{PreciseTracer's }fast response, low overhead and scalability make it a promising tracing tool for large-scale production systems.


%





\ifCLASSOPTIONcaptionsoff
  \newpage
\fi

\end{document}